\documentclass[twocolumn,aps,prl,superscriptaddress]{revtex4-2}

\usepackage{amsmath}
\usepackage{graphicx}
\usepackage{MnSymbol}
\usepackage{parskip}
\usepackage{makecell}

\begin{document}

\title{Sliding Friction of Hard Sliders on Rubber: Theory and Experiment}

\author{R. Xu}
\affiliation{State Key Laboratory of Solid Lubrication, Lanzhou Institute of Chemical Physics, Chinese Academy of Sciences, 730000 Lanzhou, China}
\affiliation{Peter Gr\"unberg Institute (PGI-1), Forschungszentrum J\"ulich, 52425, J\"ulich, Germany}
\affiliation{MultiscaleConsulting, Wolfshovener str. 2, 52428 J\"ulich, Germany}
\affiliation{Shandong Linglong Tire Co., Ltd, 265400 Zhaoyuan, China}

\author{B.N.J. Persson}
\affiliation{State Key Laboratory of Solid Lubrication, Lanzhou Institute of Chemical Physics, Chinese Academy of Sciences, 730000 Lanzhou, China}
\affiliation{Peter Gr\"unberg Institute (PGI-1), Forschungszentrum J\"ulich, 52425, J\"ulich, Germany}
\affiliation{MultiscaleConsulting, Wolfshovener str. 2, 52428 J\"ulich, Germany}
\affiliation{Shandong Linglong Tire Co., Ltd, 265400 Zhaoyuan, China}

\begin{abstract}
We present a study of sliding friction for rigid triangular steel sliders on soft rubber substrates under both lubricated and dry conditions.
For rubber surfaces lubricated with a thin film of silicone oil, the measured sliding friction at room temperature agrees well with theoretical predictions obtained from a viscoelastic model originally developed for rolling friction.
On the lubricated surface, the sliding friction is primarily due to bulk viscoelastic energy dissipation in the rubber.
The model, which includes strain-dependent softening of the rubber modulus, accurately predicts the experimental friction curves. At lower temperatures ($T = -20^\circ {\rm C}$ and $-40^\circ {\rm C}$), the measured friction exceeds the theoretical prediction.
We attribute this increase to penetration of the lubricant film by surface asperities, leading to a larger adhesive contribution.
For dry surfaces, the adhesive contribution becomes dominant.
By subtracting the viscoelastic component inferred from the lubricated case, we estimate the interfacial frictional shear stress. This shear stress increases approximately linearly with the logarithm of the sliding speed, consistent with stress-augmented thermal activation mechanisms.
\end{abstract}

\maketitle

\setcounter{page}{1}
\pagenumbering{arabic}




Corresponding author: B.N.J. Persson, 

email: b.persson@fz-juelich.de

\vskip 0.2cm

{\bf 1 Introduction} 

Rubber sliding friction plays a critical role in many applications, including tires, rubber seals, and damping systems. 
In such systems, frictional behavior is a result of the combined effects of viscoelastic energy dissipation and adhesion, which is influenced by a complex interplay between contact and environmental conditions such as surface roughness, lubrication, and temperature. 

In this study, we investigate the sliding friction of rigid triangular steel sliders on soft rubber substrates under both dry and lubricated conditions. 
This problem has been studied previously by Greenwood and Tabor \cite{GT} (see also \cite{Sabey}), who demonstrated the importance of viscoelastic contributions to friction and discussed the role of lubricant film rupture and rubber damage for sharp contacts. 
However, a detailed quantitative comparison between experimental results and theoretical predictions for viscoelastic friction, including nonlinear strain effects, has been lacking.

Here, we present new experimental results and compare them with a theoretical model originally developed for rolling friction \cite{theory}, which can be extended to sliding friction when the adhesive contribution is negligible. 
This is typically the case for lubricated interfaces where real contact areas are reduced, and fluid films suppress molecular interactions.

Our results show that, under lubricated conditions at room temperature, the viscoelastic friction model provides excellent agreement with experimental data when strain softening of the rubber is properly taken into account. 
This softening, which reflects the reduction in effective modulus at large strain amplitudes, significantly enhances the predicted friction and must be considered in quantitative modeling. 
At lower temperatures ($T = -20^\circ {\rm C}$ and $-40^\circ {\rm C}$), the measured friction exceeds the theoretical predictions. 
We attribute this to partial penetration of the lubricant film by surface asperities, leading to increased solid-solid contact and greater energy dissipation due to adhesive contributions.

For dry surfaces, adhesion contributes significantly to the total friction force. 
By subtracting the viscoelastic component obtained under lubricated conditions, we estimate the adhesive shear stress in the real contact area.
This shear stress increases approximately linearly with the logarithm of the sliding speed, consistent with thermally activated stick-slip mechanisms proposed in earlier studies.

Overall, our study demonstrates the importance of including both viscoelastic strain softening and contact mechanics in modeling sliding friction on rubber, and shows how experimental observations across different temperatures and surface conditions can be consistently interpreted within this framework.

\begin{figure}
\includegraphics[width=0.45\textwidth,angle=0.0]{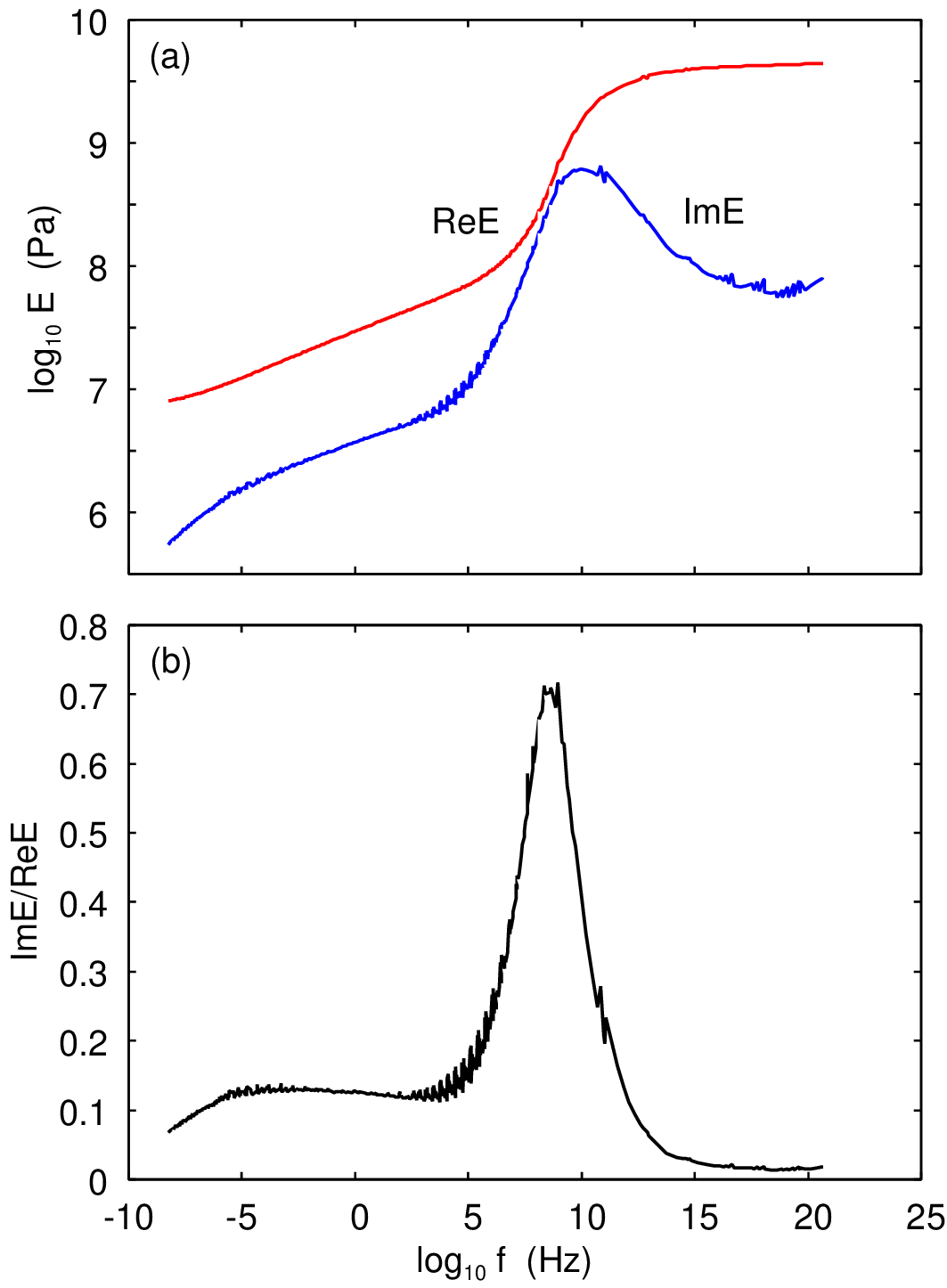}
\caption{\label{1logf.2logE.eps}
(a) The real ${\rm Re}E$ and imaginary ${\rm Im}E$ part of the linear response viscoelastic modulus, 
and (b) the tan delta ${\rm Im}E/{\rm Re}E$ as a function of the frequency for the carbon filled SB
compound for $T=20^\circ {\rm C}$. The glass transition temperature of the 
rubber is $T_{\rm g} = -50^\circ {\rm C}$.
}
\end{figure}

\begin{figure}
\includegraphics[width=0.45\textwidth,angle=0.0]{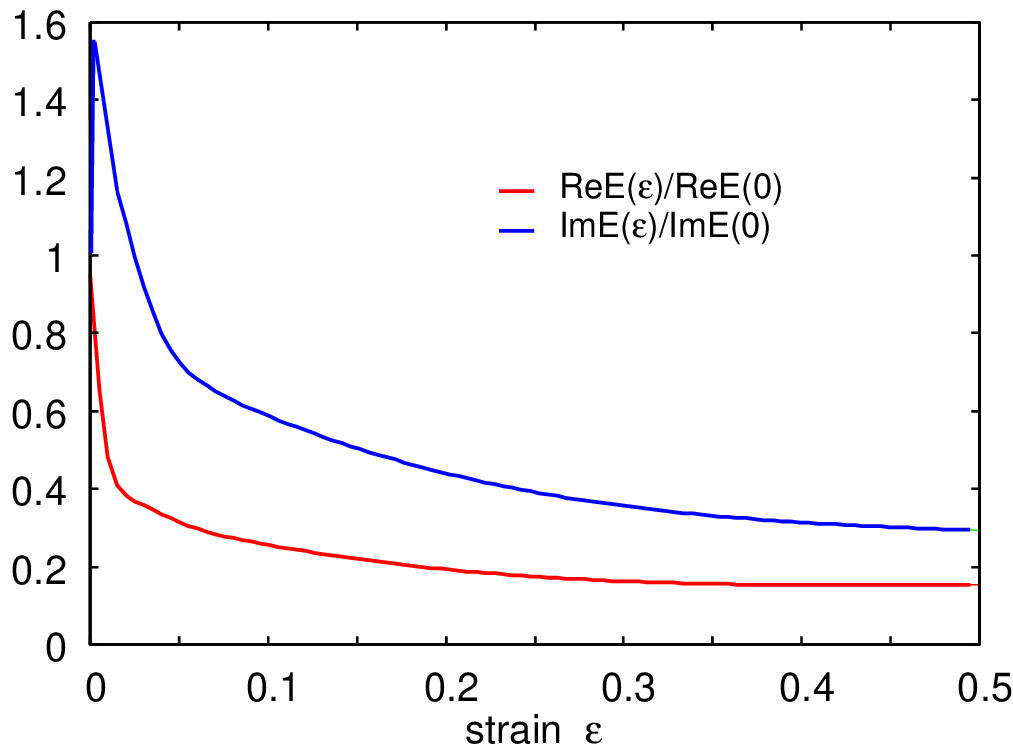}
\caption{\label{1strain.2reduction.eps}
The viscoelastic modulus at the strain $\epsilon$ divided by the modulus for
the strain $\epsilon = 10^{-4}$ obtained for the frequency $1 \ {\rm Hz}$ and the temperature
$T=20^\circ {\rm C}$.
}
\end{figure}

\vskip 0.3cm
{\bf 2 Viscoelastic modulus} 

In this study, we use the same 
carbon-filled styrene-butadiene rubber (SBR) compound as in the wear study in Ref. \cite{WEAR}.
We measured the viscoelastic modulus in elongation mode
for a strain amplitude of $4\times 10^{-4}$ (or $0.04\%$ strain) \cite{Boris}. At this small strain, we probe the
linear response properties of the rubber.
Fig. \ref{1logf.2logE.eps} shows (a) the real part ${\rm Re}E$ and the imaginary part ${\rm Im}E$, and (b) 
the ratio ${\rm Im}E / {\rm Re}E = {\rm tan}\delta$ of the viscoelastic modulus as a function of frequency.

We define the glass transition temperature $T_{\rm g}$ as the temperature 
$T$ at which ${\rm tan} \delta (T)$, 
for the frequency $\omega_{0} = 0.01 \ \rm{s}^{-1}$, reaches its maximum. 
We found that using this definition yields a $T_{\rm g}$ that is nearly the same as 
that obtained using standard methods such as viscosity or calorimetry. 
For the SBR compound used in this study, this gives 
a glass transition temperature of $T_{\rm g} \approx -50^\circ {\rm C}$.

Rubber sliding (or rolling) friction typically involves relatively large strains and depends
on the nonlinear viscoelastic properties of the rubber \cite{non1,non2,non3}.
In this paper, we studied the large-strain dependence of the effective modulus.
The measurements were performed at a frequency of $f=1 \ {\rm Hz}$ and a temperature of $T=20^\circ {\rm C}$ in elongation mode.
Fig. \ref{1strain.2reduction.eps} shows the ratio $\eta_{\rm R} (\epsilon) = {\rm Re} E(\epsilon)/{\rm Re} E(0)$ (lower curve) and
$\eta_{\rm I} (\epsilon) = {\rm Im} E(\epsilon)/{\rm Im} E(0)$ (upper curve) for the effective modulus at strain $\epsilon$ and
at zero strain (actually $\approx 10^{-4}$), as a function of $\epsilon$.

\begin{figure}[tbp]
\includegraphics[width=0.45\textwidth,angle=0.0]{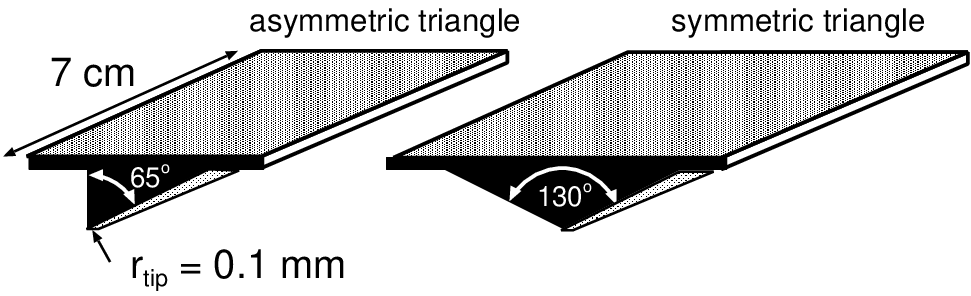}
\caption{\label{TwoTriangles.eps}
The two triangular steel sliders used in the rubber friction experiments.
}
\end{figure}

\begin{figure}[ht!]
\includegraphics[width=0.20\textwidth,angle=0.0]{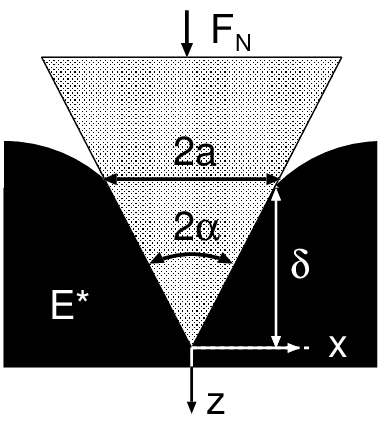}
\caption{\label{INDENTtriangle.eps}
The penetration is given by $\delta = a/{\rm tan}\alpha$. 
The length of the triangular indenter in the $y$-direction is $L_y$.
The typical strain is approximately $\delta/a = {\rm cot}\alpha$.
}
\end{figure}

\begin{figure}[tbp]
\includegraphics[width=0.45\textwidth,angle=0.0]{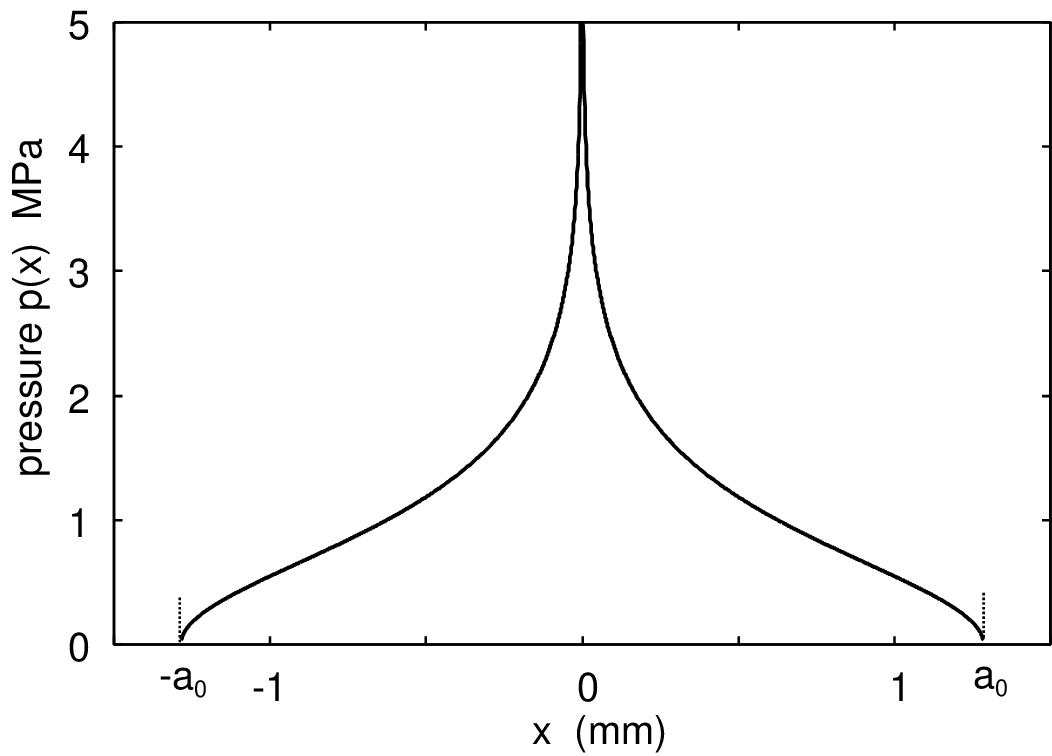}
\caption{\label{1x.2p.TRIANGLE.eps}
The pressure $p(x)$ as a function of $x$ for the symmetric indenter pressed
against the rubber with a normal force per unit length $F_{\rm N}/L_y = 3000 \ {\rm N/m}$.
The effective elastic modulus, including a strain softening factor of $0.2$, is taken as 
$E_{\rm eff} (1/t_0,\epsilon) = 25 \times 0.2 = 5 \ {\rm MPa}$, corresponding to a contact time of approximately
$t_0 \approx 1000 \ {\rm s}$. The indenter tip is assumed to be perfectly sharp, $r_{\rm tip}=0$.
The width of the contact region is $w=2a_0 \approx 2.5 \ {\rm mm}$.
}
\end{figure}

\vspace{0.3cm}
{\bf 3 Theory of viscoelastic contribution to rubber sliding friction}

We consider lubricated friction at low sliding speed and assume that the frictional
shear stress in the area of real contact is negligible. 
The experimental results will be compared to the theory 
developed in Ref. \cite{theory} for rolling friction,
which is also valid for lubricated sliding friction at low speed if 
the frictional shear stress is negligible. 

The theory presented in Ref. \cite{theory} predicts 
the rolling friction for rigid spheres and cylinders on flat surfaces of viscoelastic solids.
It was found to agree very well with
the exact result of Hunter for rolling cylinders, 
for the simple rheological model used by Hunter \cite{Hunt}. 
Here, we apply the same theory to triangular steel profiles sliding on a lubricated rubber plate.

In this study, we use the triangular steel profiles shown in Fig. \ref{TwoTriangles.eps}.
A profile is pressed against a flat rubber sheet (thickness $d$) with the
normal force per unit length $f_{\rm N} = F_{\rm N} /L_y$, 
and slides at speed $v$. If the width $2a$ (in the sliding direction)
of the footprint of the triangle and the rubber sheet 
satisfy $2a \ll d$, we can treat the rubber slab as infinitely thick. 
In this case, the sliding friction force \cite{theory} is given by
$$F_{\rm f} = {2 (2\pi )^2 \over v} \int d^2q \ {\omega \over q} {\rm Im} \left[{1\over E_{\rm eff} (\omega,\epsilon)}\right] |p({\bf q})|^2 \eqno(1)$$
where $E_{\rm eff}$ is the
effective viscoelastic modulus defined as
${\rm Re}E_{\rm eff} (\omega, \epsilon)= \eta_{\rm R} (\epsilon) {\rm Re}E^*(\omega)$
and ${\rm Im}E_{\rm eff} (\omega, \epsilon) = \eta_{\rm I} (\epsilon) {\rm Im}E^*(\omega)$, 
with $E^*(\omega) = E(\omega) /(1-\nu^2)$. 
Here, $\epsilon$ is the typical strain in the contact region (see below).
The integration variable $q$ represents the wavenumber and $\omega = {\bf q} \cdot {\bf v}$. In (1), $p({\bf q})$ is the
Fourier transform of the contact pressure $p({\bf x})$.

We consider a symmetric triangular indenter with a half-opening angle $\alpha$ (see Fig. \ref{INDENTtriangle.eps}) 
pressed against an elastic half-space with a force per unit length $f_{\rm N} = F_{\rm N}/L_y$. The
pressure distribution \cite{Johnson} is given by
$$p(x)= p_0 {\rm cosh}^{-1} \left ({a\over x}\right )\eqno(2)$$

To determine the prefactor $p_0$, we note that
$$f_{\rm N} = \int_{-a}^a dx \ p_0 {\rm cosh}^{-1} \left ( {a\over x} \right )
= a p_0 \int_{-1}^1 d\xi \ {\rm cosh}^{-1} \left ( {1\over \xi} \right )$$ 
Since ${\rm cosh}^{-1}$ is even and defined on $[1,\infty)$, we can write the integral as
$$f_{\rm N} = 2 a p_0 \int_0^1 d\xi \ {\rm cosh}^{-1} \left ( {1\over \xi} \right )=\pi a p_0$$
thus,
$$p_0 = { f_{\rm N} \over \pi a}$$
The half-width of the contact is
$$a = { f_{\rm N} \over E^* {\rm cot}\alpha }\eqno(3)$$
where $E^* = E/(1-\nu^2)$.
When used in (1), $E^*$ is replaced with $|E_{\rm eff} (\omega ,\epsilon)|$, 
with $\omega = {\bf q}\cdot {\bf v}$, so the contact 
width $w=2a$ depends on the sliding speed $v$. Fig. \ref{1x.2p.TRIANGLE.eps}
shows the pressure $p(x)$ as a function of $x$ for the symmetric indenter squeezed
against the rubber with a normal force per unit length $F_{\rm N}/L_y = 3000 \ {\rm N/m}$.
We use the effective elastic modulus corresponding to a contact time of the order of
$\sim 1000 \ {\rm s}$. The pressure distribution exhibits a (weak) singularity 
$p\sim \log|x|$ as $x\rightarrow 0$. 


The average pressure is given by
$$\bar p = {1\over 2 a} \int_{-a}^a dx \ p(x) = {\pi \over 2} p_0$$
or $\bar p =  f_{\rm N}/2 a$. The penetration $\delta$ is given by (see Fig. \ref{INDENTtriangle.eps}):
$$\delta = {a\over {\rm tan}\alpha}$$

The Fourier transform of the pressure is
$$p ({\bf q}) = {1\over (2\pi )^2} \int d^2 x \ p ({\bf x}) e^{-i{\bf q}\cdot {\bf x}} = \delta (q_y) p(q_x)$$
where
$$p(q_x) = {1\over 2 \pi} \int dx \ p (x) e^{-iq_x x}$$
$$= {p_0 \over 2 \pi} \int_{-a}^a dx \ {\rm cosh}^{-1} \left ( {a\over x} \right ) e^{-iq_x x}$$
$$= {p_0 a \over 2 \pi} \int_{-1}^1 d\xi \ {\rm cosh}^{-1} \left ( {1 \over \xi} \right )  e^{-iq_x a \xi}$$
Define
$$g(q_x a) = {1\over 2 \pi} \int_{-1}^1 d\xi \ {\rm cosh}^{-1} \left ( {1 \over \xi} \right )  e^{-iq_x a \xi}$$
$$= {1\over  \pi} \int_0^1 d\xi \ {\rm cosh}^{-1} \left ( {1 \over \xi} \right )  {\rm cos}( q_x a \xi)$$
we obtain
$$p(q_x) = {f_{\rm N} \over \pi} g(q_x a)\eqno(4)$$ 
Using
$$[\delta (q_y)]^2 = \delta (q_y) \cdot {1\over 2 \pi} \int dy \ e^{-iq_y y } = {L_y \over 2 \pi} \delta (q_y)$$
we get from (1):
$$\mu = {F_{\rm f}\over f_{\rm N} L_y} = {8 \pi \over f_{\rm N}} 
\int_0^\infty dq_x \ {\rm Im} \left[{1\over E_{\rm eff} (q_x v,\epsilon)}\right] |p_0 (q_x)|^2$$
Substituting (4) gives:
$$\mu = {8 f_{\rm N} \over \pi} \int_0^\infty dq_x \ {\rm Im} \left[{1\over E_{\rm eff} (q_x v, \epsilon)}\right] |g (q_x a)|^2\eqno(5)$$
The triangular steel profiles have a finite tip radius of curvature $r_{\rm tip} = 0.1 \ {\rm mm}$. We account for this
by using a high-wavenumber cutoff in the integral (5), limiting $0 < q_x < q_1$ with $q_1 = 2 \pi /r_{\rm tip}$.

The strain in the region where the viscoelastic deformations occur varies spatially,
but is typically of the order
$$\epsilon = {1\over 2} \left ({\partial u_z \over \partial x} + {\partial u_x \over \partial z} \right ) 
\approx  {1\over 2} {\partial u_z \over \partial x} \approx {\delta \over 2 a} \approx {1 \over 2 {\rm tan}\alpha} . \eqno(6)$$
As the sliding speed increases (or the temperature decreases),
the effective modulus $E_{\rm eff} (qv,\epsilon)$ increases and the half-width $a$
decreases, while the strain $\epsilon$ remains unchanged. In our case, $\epsilon \approx 0.25$.
In the analysis below, we present theoretical results that include
strain softening. These were obtained using the effective 
modulus $E_{\rm eff} (\omega,\epsilon)$
for the strain $\epsilon$ given by (6). 
This approach is, of course, approximate, since it is based on linear
viscoelastic theory with a correction for nonlinearity introduced via an effective modulus
that depends on the average strain in the contact region, while the actual strain varies
in space. Still, comparison with experimental data (see below)
shows that this approach provides a good estimation of the influence
of strain softening on sliding friction.

For the strain $\epsilon \approx 0.25$, based on Fig. \ref{1strain.2reduction.eps},
we obtain $\eta_{\rm R} \approx 0.2$ and $\eta_{\rm I} \approx 0.4$.
Since sliding friction depends on ${\rm Im} [1/E_{\rm eff}]$, 
the nonlinear (strain softening) correction will be significant,
enhancing the sliding friction by roughly a factor of 5 (see below),
which is similar to what was found in earlier rolling friction studies.

\begin{figure}[tbp]
\includegraphics[width=0.45\textwidth,angle=0.0]{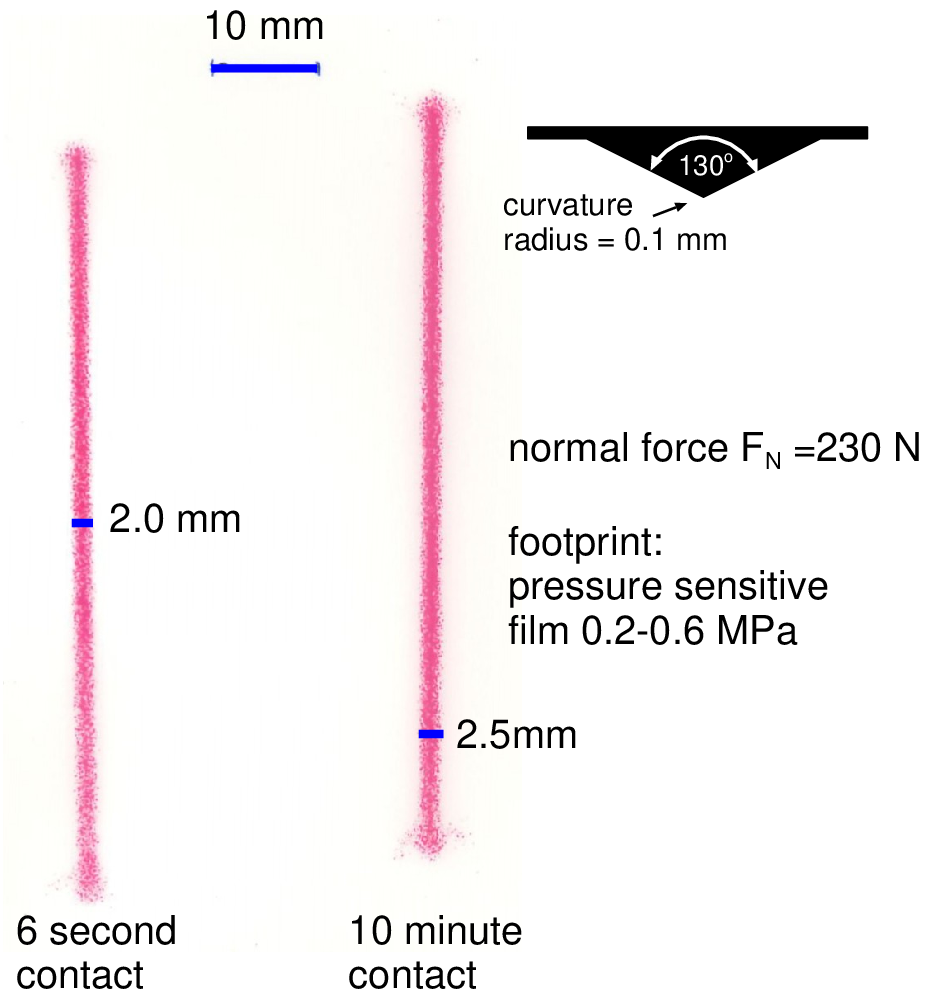}
\caption{\label{triangle.eps}
The metal triangle loaded against the rubber surface with a pressure-sensitive film in between.
The two footprints correspond to contact times of $6 \ {\rm s}$ and $600 \ {\rm s}$, respectively.
The two short blue bars indicate widths of $2$ and $2.5 \ {\rm mm}$.
}
\end{figure}

\vspace{0.3cm}
{\bf 3 Experimental results and comparison to theory}

We used the same linear friction tester as in previous studies on sliding and rolling friction \cite{rolling}.
Two different triangular steel profiles (see Fig. \ref{TwoTriangles.eps}) were 
attached to the force cell.  
The steel surfaces exhibit self-affine fractal roughness with a Hurst exponent $H \approx 0.77$, 
a roll-off wavenumber $q_{\rm r} \approx 2\times 10^4 \ {\rm m}^{-1}$, 
and a root-mean-square roughness $h_{\rm rms} = 1.5 \ {\rm \mu m}$.
A SBR plate of thickness $15 \ {\rm mm}$ 
was fixed to the machine table, which was driven by a servo motor through a gearbox to produce translational motion. 
The relative velocity between the steel triangle and the rubber substrate was controlled, 
while the force cell recorded both the normal force and the friction force.
The normal load per unit length was $F_{\rm N}/L_y = 3000 \ {\rm N/m}$, and the sliding 
speed was varied from $1 \ {\rm \mu m/s}$ to $1 \ {\rm cm/s}$. 
Both triangular steel sliders were $7 \ {\rm cm}$ long in the $y$-direction, orthogonal to the sliding direction.
The tip angles were $65^\circ$ and $130^\circ$, and the tips have a radius of curvature $r_{\rm tip} = 0.1 \ {\rm mm}$.

In Fig. \ref{triangle.eps}, we show two pressure footprints of the contact between the symmetric triangle slider
and the rubber substrate obtained using a pressure-sensitive film. 
The metal triangle was loaded against the rubber surface with the film in between.
The two footprints correspond to contact times of $6 \ {\rm s}$ and $600 \ {\rm s}$, 
with widths (indicated by the blue bars) of $2$ and $2.5 \ {\rm mm}$, respectively.

\begin{figure}[tbp]
\includegraphics[width=0.45\textwidth,angle=0.0]{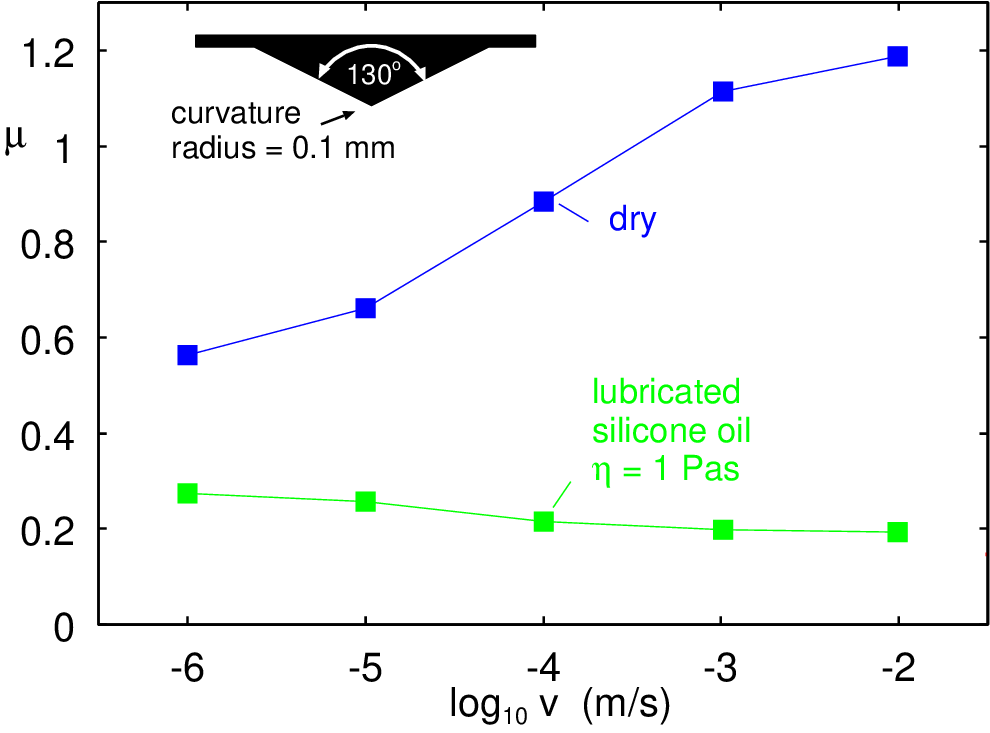}
\caption{\label{logv.2mu.WithExp1.eps}
Sliding friction for the symmetric triangle on dry (blue squares) and lubricated (green squares) rubber.
}
\end{figure}

Fig. \ref{logv.2mu.WithExp1.eps}
shows the measured sliding friction for the symmetric triangle on dry (blue squares) and lubricated (green squares) rubber.
The lubricated surface was prepared by spreading a teaspoon of silicone oil 
(DuPont Liveo 360 medical fluid, viscosity $0.97 \ {\rm Pa\cdot s}$) over the rubber surface area ($14 \ {\rm cm} \times 14 \ {\rm cm}$).
Due to the low sliding speed and high contact pressure, the silicone oil acts as a boundary lubricant
(without hydrodynamic effects) and likely eliminates or reduces any opening-crack contribution to the sliding
friction (see below).

In Fig. \ref{1logv.2mu.three.radius.and.exp.eps}, we compare the measured friction for the lubricated surface 
with the theoretical prediction at room temperature. The green squares are the experimental data 
from Fig. \ref{logv.2mu.WithExp1.eps}, and the red line is the theoretical prediction 
of the viscoelastic contribution to friction for the symmetric triangle.
The theory involves no fitting parameters, and the agreement between theory and experiment is remarkably good.
This result shows that, for the lubricated surface at $T = 20^\circ {\rm C}$, 
the measured friction is almost entirely due to the viscoelastic
deformation of the substrate. This implies that the difference between dry and lubricated friction in
Fig. \ref{logv.2mu.WithExp1.eps} must result from the adhesive contribution in the area of real contact.
 
The red curve in Fig. \ref{1logv.2mu.three.radius.and.exp.eps} corresponds to the actual tip radius of $0.1 \ {\rm mm}$.
The pink and blue curves represent theoretical results for
$r_{\rm tip} = 0.2$ and $0.05 \ {\rm mm}$, respectively.
As expected, a sharper tip leads to slightly higher sliding friction, but the effect is small due to the 
weak (logarithmic) singularity of the contact pressure near the tip at $x = 0$ 
as $r_{\rm tip} \rightarrow 0$.

\begin{figure}[tbp]
\includegraphics[width=0.45\textwidth,angle=0.0]{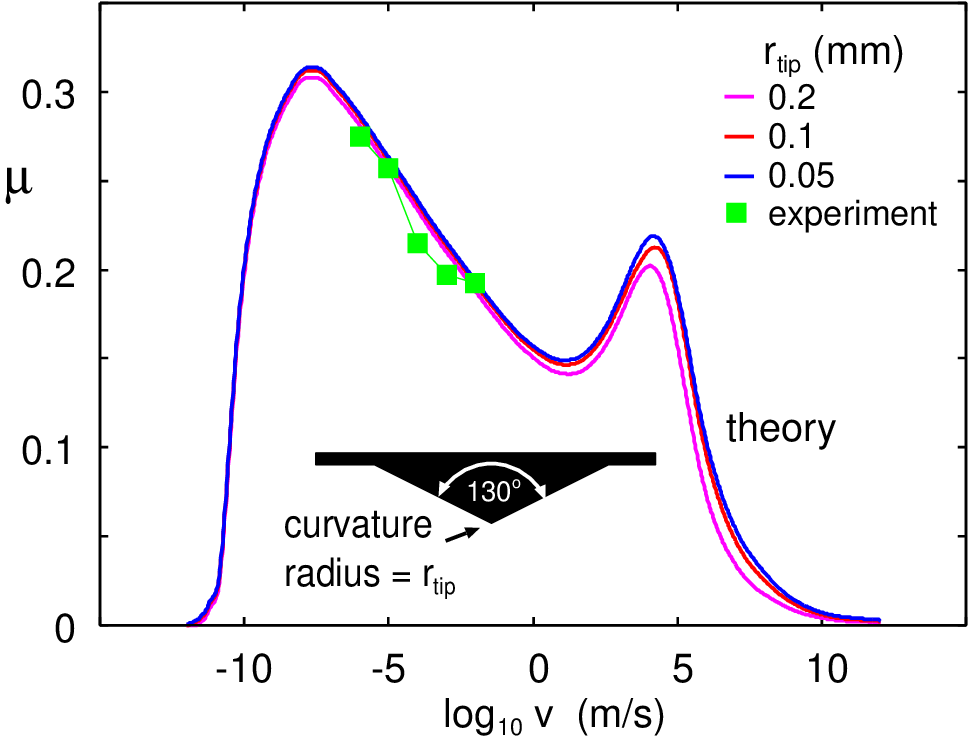}
\caption{\label{1logv.2mu.three.radius.and.exp.eps}
Sliding friction for the symmetric triangle on the lubricated rubber surface (green squares from Fig. \ref{logv.2mu.WithExp1.eps}).
The red line is the theoretically predicted viscoelastic contribution to the friction, 
assuming the measured tip radius $r = 0.1 \ {\rm mm}$.
The pink and blue lines show the predicted values for tip radii of $0.2$ and $0.05 \ {\rm mm}$, respectively.
}
\end{figure}

\begin{figure}[tbp]
\includegraphics[width=0.45\textwidth,angle=0.0]{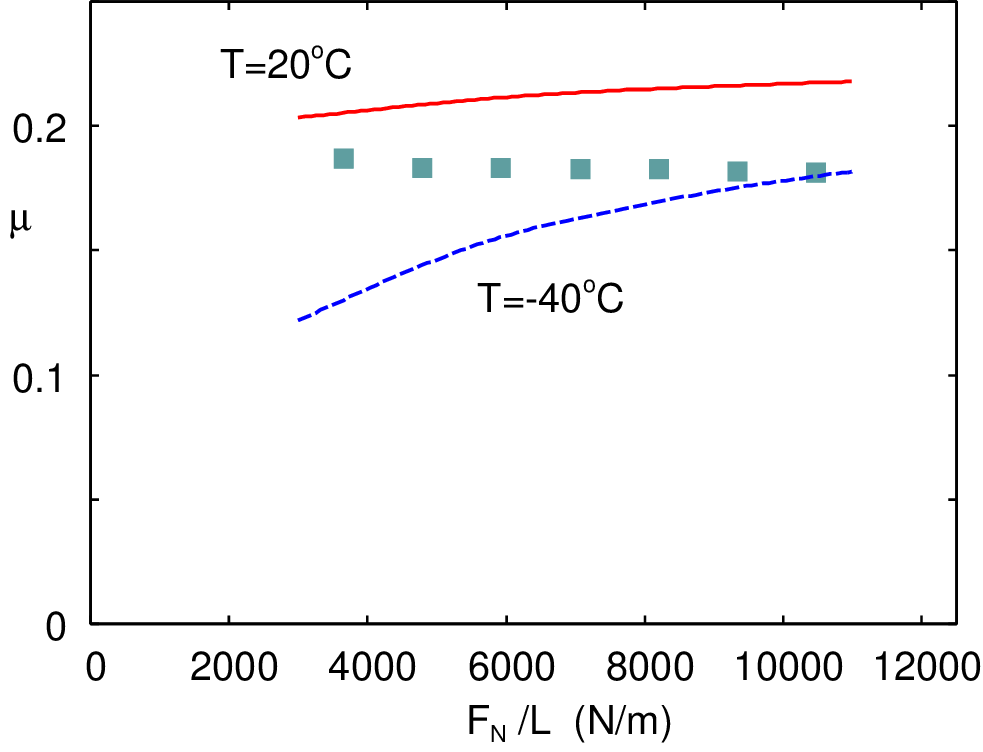}
\caption{\label{1pressure.2mu.eps}
Dependence of the friction coefficient on the normal load 
for a symmetric slider sliding on a lubricated rubber substrate. The measured data (squares) and the red line represent the experimental and theoretical results at $T = 20^\circ {\rm C}$.
The blue dashed line corresponds to the calculated result at $T = -40^\circ {\rm C}$.
Theoretical predictions of the viscoelastic contributions were obtained using the measured tip radius of $r = 0.1 \ {\rm mm}$.
}
\end{figure}

\begin{figure}[tbp]
\includegraphics[width=0.45\textwidth,angle=0.0]{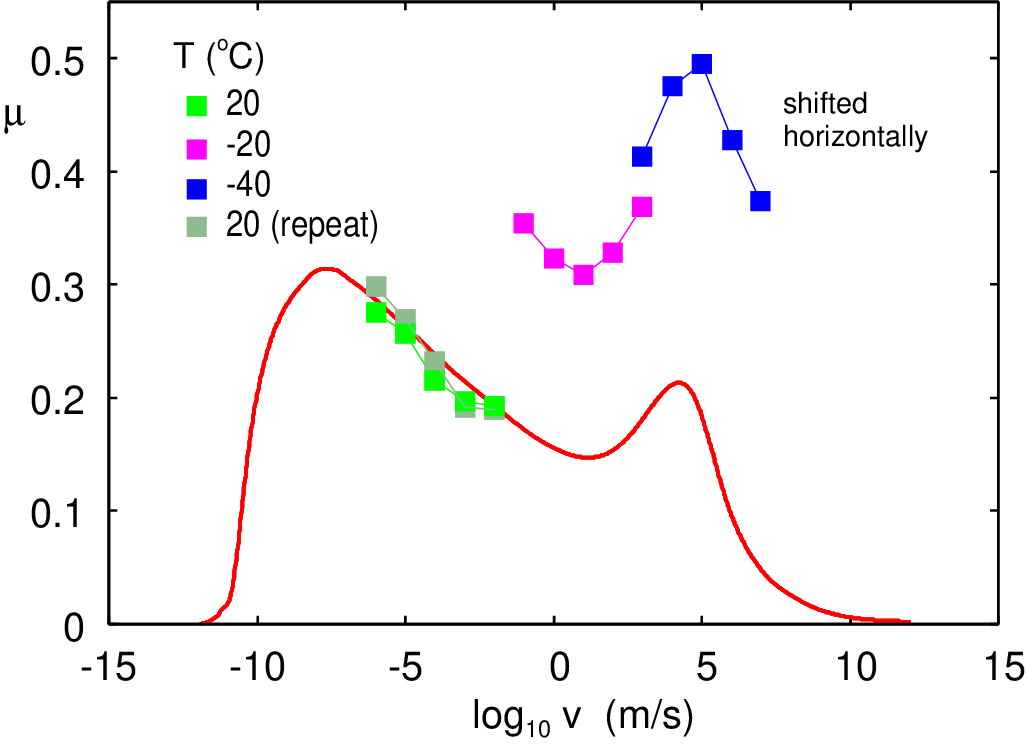}
\caption{\label{1logv.2mu.all.temp.shift1.eps}
Sliding friction for the symmetric triangle on lubricated rubber at $T = 20$ (green), $-20$ (pink), and $-40^\circ {\rm C}$ (blue), 
along with a repeated measurement at $20^\circ {\rm C}$ (dark green). 
The measured data for $-20^\circ {\rm C}$ and $-40^\circ {\rm C}$ are shifted horizontally using the viscoelastic bulk shift factor.
The red line is the theoretically predicted viscoelastic contribution assuming a tip radius of $r = 0.1 \ {\rm mm}$.
}
\end{figure}

Fig. \ref{1pressure.2mu.eps} shows the dependence of the friction coefficient on the normal load for a symmetric slider sliding on a lubricated rubber substrate. The sliding velocity was $v = 3 \ {\rm mm/s}$. The measured data (squares) and the red line represent the experimental and theoretical results, respectively, at $T = 20^\circ {\rm C}$. The blue dashed line corresponds to the calculated result at $T = -40^\circ {\rm C}$. Theoretical predictions of the viscoelastic contributions were obtained using the measured tip radius of $r = 0.1 \ {\rm mm}$. It can be seen that the friction coefficient in the theoretical prediction increases slightly with increasing load, whereas the experimental results exhibit the opposite trend and show a slight decrease in the friction coefficient with increasing load.

Fig. \ref{1logv.2mu.all.temp.shift1.eps} shows the measured sliding friction for the symmetric slider on the lubricated rubber substrate at different temperatures: $T = 20^\circ {\rm C}$ (green data points, from Fig.~\ref{1logv.2mu.three.radius.and.exp.eps}), $-20^\circ {\rm C}$ (pink), $-40^\circ {\rm C}$ (blue), and a repeat measurement at $20^\circ {\rm C}$ (dark green).
The red line represents the theoretically predicted viscoelastic contribution to the friction for the symmetric slider, based on the measured tip radius $r = 0.1 \ {\rm mm}$. The measured data at $-20^\circ {\rm C}$ and $-40^\circ {\rm C}$ are shifted horizontally using the viscoelastic bulk shift factor, so that they correspond to the equivalent friction values at $20^\circ {\rm C}$ but at higher sliding speeds.
However, these measured friction values lie significantly above the theoretical prediction (red curve).

\begin{figure}[tbp]
\includegraphics[width=0.45\textwidth,angle=0.0]{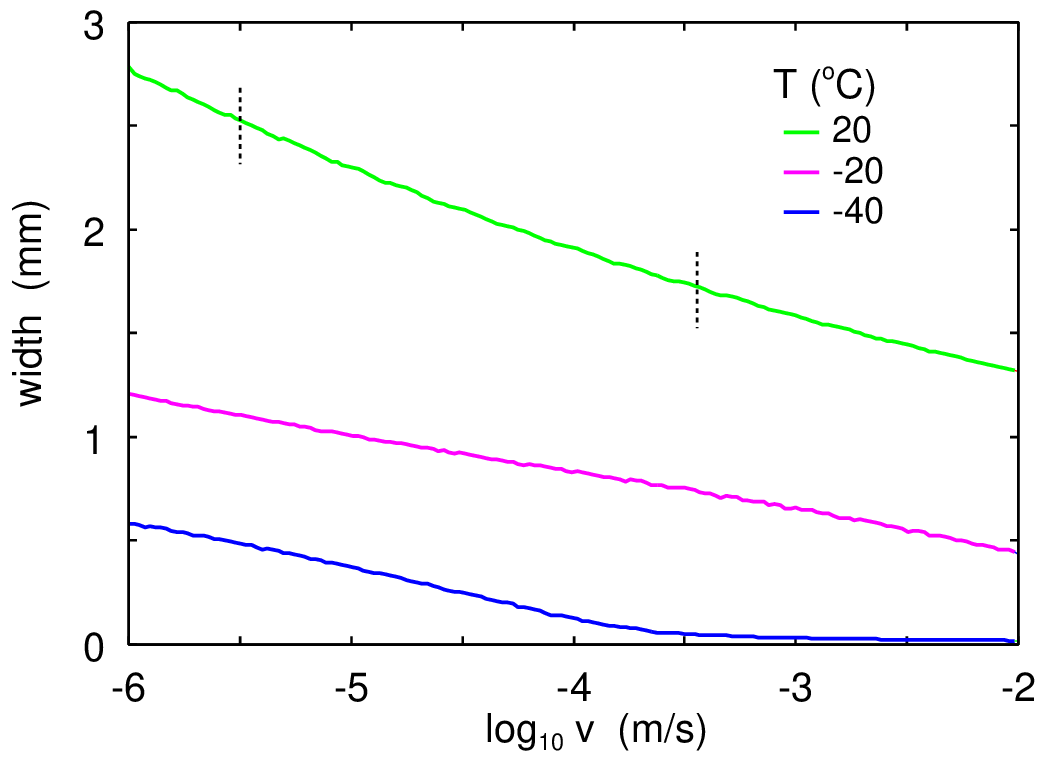}
\caption{\label{1logv.2width.all.temperatures.eps}
The calculated contact width as a function of sliding speed,
for temperatures $T = 20^\circ {\rm C}$, $-20^\circ {\rm C}$, and $-40^\circ {\rm C}$.
}
\end{figure}

We attribute the enhanced friction at $-20^\circ {\rm C}$ and $-40^\circ {\rm C}$ to the high contact pressure resulting from the increased elastic modulus of rubber at low temperatures, which may lead to penetration of the lubricant film by surface asperities. To support this interpretation, Fig. \ref{1logv.2width.all.temperatures.eps} shows the calculated width of the contact region as a function of sliding speed for $T = 20^\circ {\rm C}$, $-20^\circ {\rm C}$, and $-40^\circ {\rm C}$.
The contact width is much smaller at the two lower temperatures, which corresponds to a higher contact pressure.

Nevertheless, the sliding friction at $-20^\circ {\rm C}$ and $-40^\circ {\rm C}$ remains lower than expected for dry surfaces, indicating that the lubricant film continues to reduce friction even under these colder conditions.

Assuming, as supported by the theory, that the sliding friction on the lubricated surface 
at room temperature is dominated by viscoelastic effects, we can estimate the shear stress 
acting in the area of real contact under dry conditions using
$$\tau_{\rm f} (v) = [\mu (v)-\mu_{\rm visc} (v)] {F_{\rm N} \over w(v)L_y}$$
Here, $\mu(v)$ is the total friction coefficient, and
$\mu_{\rm visc}$ is the viscoelastic contribution, as given by the green and blue curves
in Fig. \ref{logv.2mu.WithExp1.eps}.
The normal force is $F_{\rm N} \approx 210 \ {\rm N}$, and the real contact area is 
$A = w L_y$, where $w(v)$ is the width of the contact and $L_y = 7 \ {\rm cm}$ is the slider length. 
Note that $\tau_{\rm f}(v)$ increases nearly linearly with the logarithm of the sliding speed,
as expected for thermally activated processes in the low-velocity regime.

\begin{figure}[tbp]
\includegraphics[width=0.45\textwidth,angle=0.0]{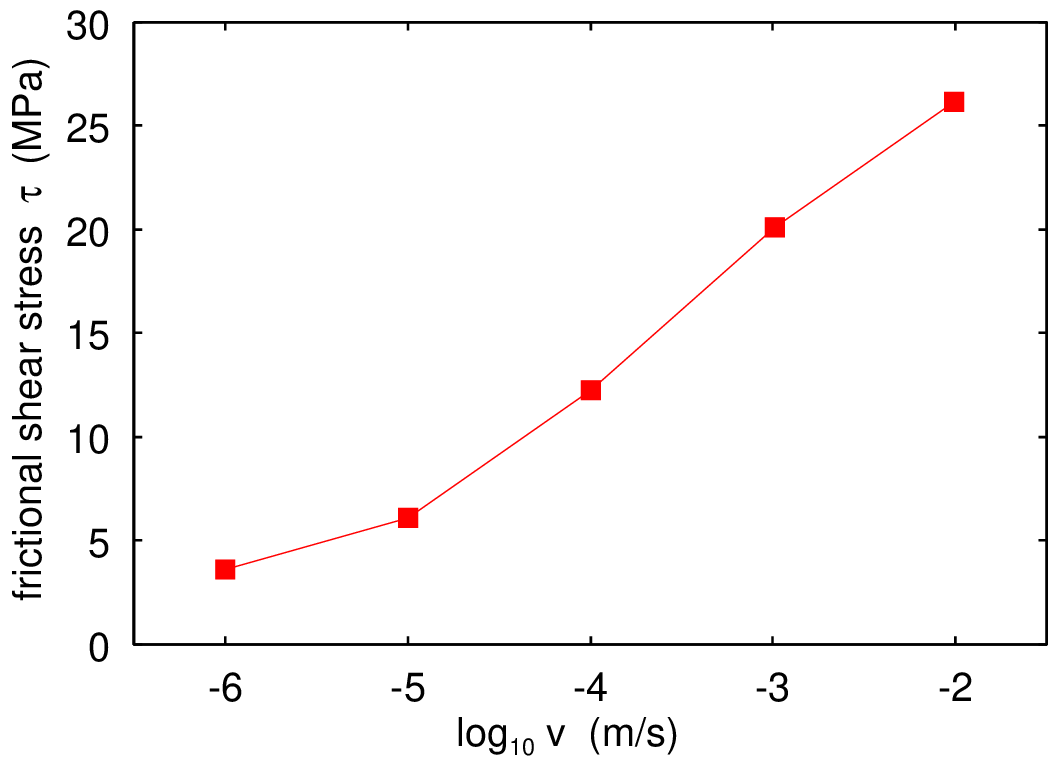}
\caption{\label{1logv.2ShearStress.eps}
The frictional shear stress in the area of real contact, obtained by subtracting the viscoelastic contribution 
$\mu_{\rm visc}(v)$ from the total friction coefficient $\mu(v)$ (green and blue curves in Fig. \ref{logv.2mu.WithExp1.eps}),
and multiplying the difference by the normal force $F_{\rm N} \approx 210 \ {\rm N}$. 
The result is divided by the contact area $A = w(v) L_y$, where $w(v)$ is the contact width and $L_y = 7 \ {\rm cm}$:
$\tau (v) = [\mu (v)-\mu_{\rm visc} (v)] F_{\rm N}/w(v)L_y$.
}
\end{figure}

Fig. \ref{1logv.2width.all.temperatures.eps} shows the calculated width of the contact region as a function of the sliding speed.
Note that at room temperature (green curve), the predicted widths are very similar to the contact width of $2$–$2.5 \ {\rm mm}$ obtained using pressure-sensitive film (see Fig. \ref{triangle.eps}).
In fact, the results in Fig. \ref{1logv.2width.all.temperatures.eps} can be used to accurately estimate the contact width after $6 \ {\rm s}$ and $600 \ {\rm s}$ of stationary contact as follows.

During a time interval $t$, the slider moves a distance $d = vt$, so the contact width $d = 2 \ {\rm mm}$ observed after $t = 6 \ {\rm s}$ corresponds to a sliding speed of $v \approx d/t \approx 3 \times 10^{-4} \ {\rm m/s}$.
Similarly, a contact width of $d = 2.5 \ {\rm mm}$ for $t = 600 \ {\rm s}$ corresponds to a sliding speed of $v = 4.2 \times 10^{-6} \ {\rm m/s}$.
These two speeds are indicated by vertical dotted lines in Fig. \ref{1logv.2width.all.temperatures.eps}, the corresponding contact widths are approximately $w \approx 1.8 \ {\rm mm}$ and $2.4 \ {\rm mm}$, respectively, in nearly perfect agreement with the measured footprint widths shown in Fig. \ref{triangle.eps}.

It is important to note that this good agreement with the theory is obtained only when strain softening is included.
Since the contact width $w = 2a$ is proportional to $1/E^*$ [see Eq. (3)], using the low-strain modulus would result in a predicted width approximately five times smaller than observed.

\begin{figure}[tbp]
\includegraphics[width=0.45\textwidth,angle=0.0]{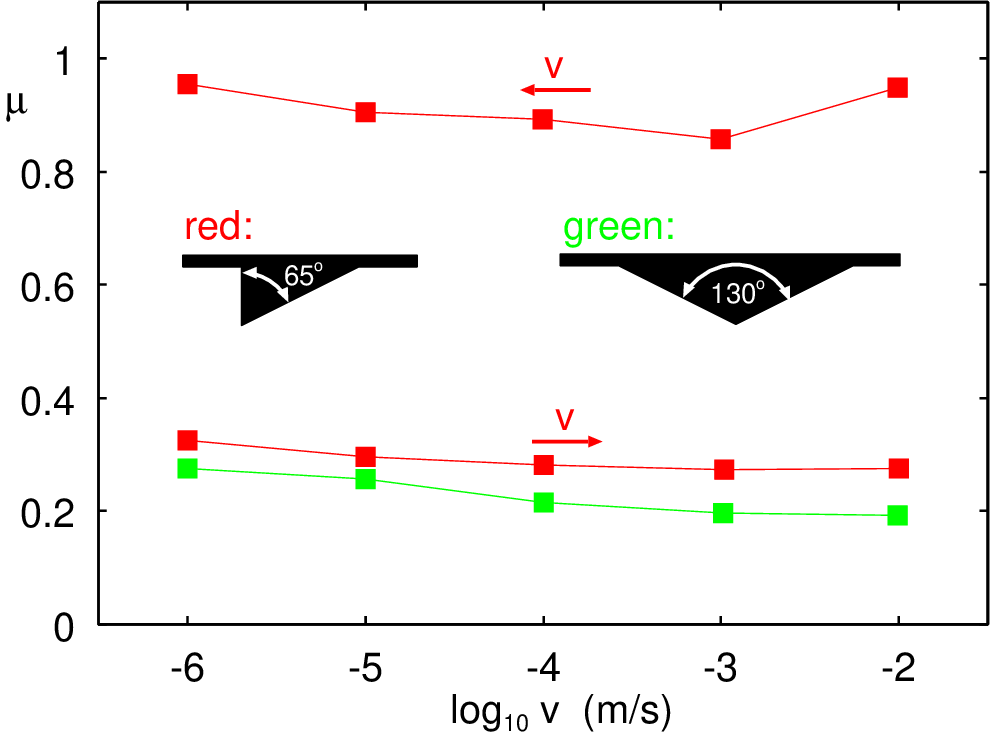}
\caption{\label{TwoTriangles.1logv.2mu.eps}
Sliding friction for two triangular sliders on a lubricated rubber substrate.
}
\end{figure}

\begin{figure}[tbp]
\includegraphics[width=0.45\textwidth,angle=0.0]{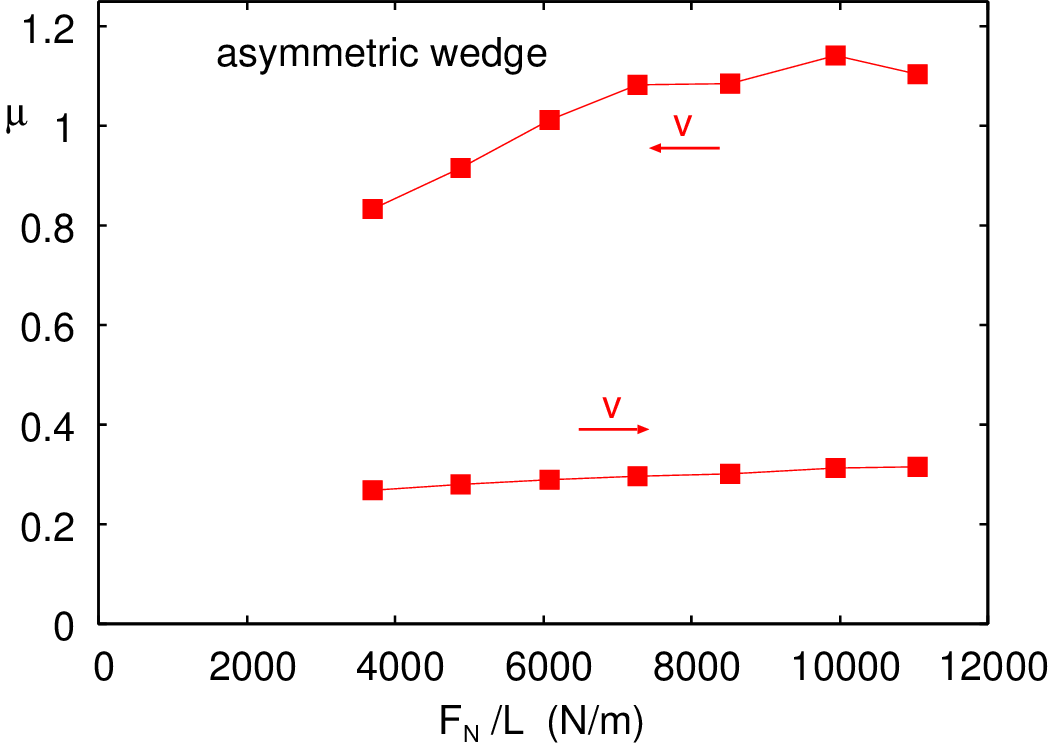}
\caption{\label{1pressure.2mu.asymetricREPEAT.eps}
Dependence of the friction coefficient on the normal load 
for an asymmetric slider sliding in the forward and backward directions on a lubricated rubber substrate.
}
\end{figure}

\begin{figure}[tbp]
\includegraphics[width=0.45\textwidth,angle=0.0]{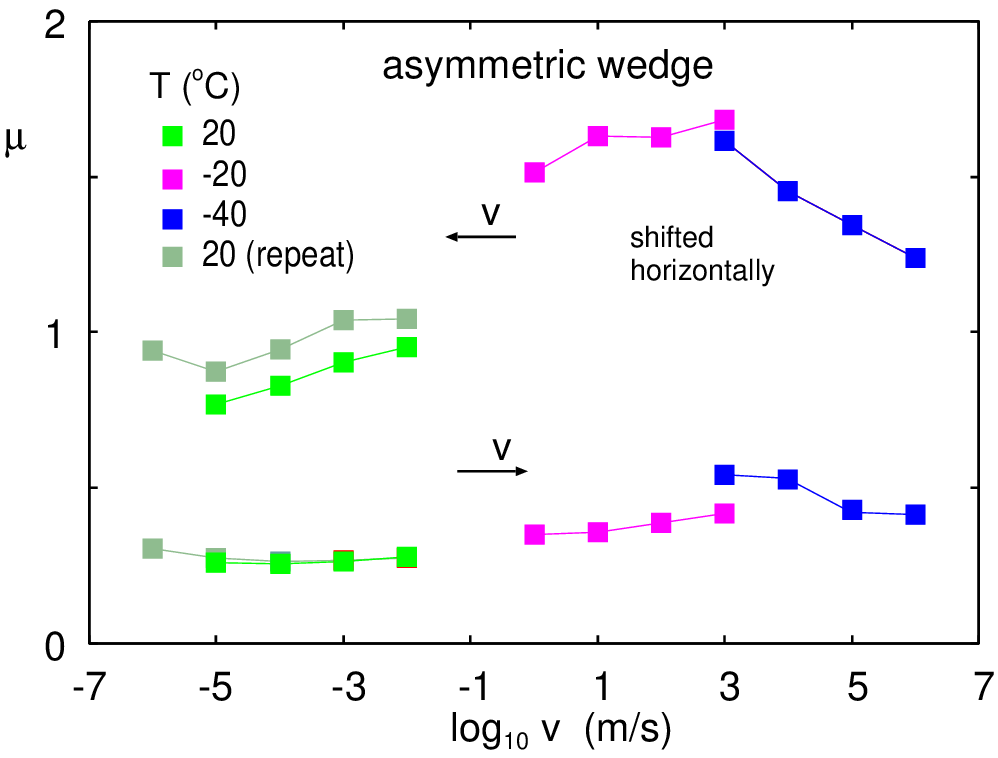}
\caption{\label{XX.1logv.2mu.asymmetric.eps}
Sliding friction for the asymmetric slider on the lubricated rubber at $T = 20^\circ {\rm C}$ (green), $-20^\circ {\rm C}$ (pink), and $-40^\circ {\rm C}$ (blue), along with a repeat measurement at $20^\circ {\rm C}$ (dark green).
The measured data for $-20^\circ {\rm C}$ and $-40^\circ {\rm C}$ have been shifted horizontally using the bulk viscoelastic shift factor.
}
\end{figure}

We have also performed sliding friction experiments for the asymmetric slider, but only under lubricated conditions.
In this case, the friction differs significantly depending on the sliding direction.
Fig. \ref{TwoTriangles.1logv.2mu.eps} shows that when sliding in the direction of the slope (forward), the friction is only slightly higher than that measured for the symmetric slider.
When sliding in the direction of the vertical side of the steel sliders (backward), the friction is very high, approximately equal to 1, even on the lubricated surface.
In this configuration, the edge of the steel profile likely penetrates the lubricant film, resulting in the high observed friction coefficient.

Fig. \ref{1pressure.2mu.asymetricREPEAT.eps} shows the dependence of the friction coefficient on the normal load 
for the asymmetric slider sliding in the forward and backward directions on the lubricated rubber substrate.
Notably, in the forward sliding direction, the friction coefficient increases slightly with increasing normal load.
This trend contrasts with that observed for the symmetric slider, where the friction coefficient decreases slightly as the load increases, as shown in Fig. \ref{1pressure.2mu.eps}.

Fig. \ref{XX.1logv.2mu.asymmetric.eps} presents the sliding friction for the asymmetric slider on the lubricated rubber at $T = 20^\circ {\rm C}$ (green), $-20^\circ {\rm C}$ (pink), and $-40^\circ {\rm C}$ (blue), along with a repeat measurement at $20^\circ {\rm C}$ (dark green).
The data for $-20^\circ {\rm C}$ and $-40^\circ {\rm C}$ have been shifted horizontally using the bulk viscoelastic shift factor.
At these lower temperatures, the measured friction in the forward sliding direction is comparable in magnitude to that of the symmetric slider.

For the asymmetric slider, we do not have numerical results for the viscoelastic contribution to friction.
However, it is likely that, as in the case of the symmetric slider, the theoretical viscoelastic contribution is much smaller than the measured values.
This discrepancy may result from penetration of the lubricant film at low temperatures, where the rubber is significantly stiffer than at room temperature and, as a result, the contact pressure is much higher.

\vspace{0.3cm}
{\bf 4 Discussion}

Greenwood and Tabor \cite{GT} found that the rolling friction coefficient for a steel sphere on a rubber surface
was nearly the same as the sliding friction for the same sphere on the same rubber surface lubricated
with a wet soap film.
This indicates that, at least under the conditions of their study, the viscoelastic contribution dominates the friction.

They also studied the friction of cones on the same lubricated rubber surface and argued that,
for large tip angles, the sliding friction is primarily viscoelastic.
However, as the cone opening angle decreased, they observed a significant increase in friction,
which they attributed to penetration of the lubricant film and tearing of the rubber.
Specifically, cones with half-angles of $80^\circ$, $60^\circ$, and $50^\circ$ produced practically no damage to the rubber
and exhibited relatively low friction coefficients in the range of $0.1$ to $0.2$.
In contrast, a $45^\circ$ cone caused light wear, while a $30^\circ$ cone produced substantial wear along the center of the contact region.

In our experiments, we observed no visible wear, even for the asymmetric slider moving on the unlubricated rubber surface
in the direction of the vertical wall, where the friction coefficient was similarly high (on the order of 1)
as that reported in Ref. \cite{GT} for the $30^\circ$ cone.
This difference may be due to the distinct wear properties of the rubber compounds used,
although no information about the rubber type was provided in Ref. \cite{GT}.

In a recent study, Ciavarella et al. \cite{Cia} used the boundary element method 
to investigate the sliding friction of non-cylindrical punches with power-law profiles $|x|^n$
on a model ``rubber'' characterized by a single relaxation time.
They observed good agreement with the prediction of Eq. (1) for $n = 2$ (case for Hunter \cite{Hunt}),
but some deviations were found for larger values of $n$.
However, they showed that if the solution for the cylindrical case ($n = 2$) was normalized 
using the modulus and mean pressure at zero sliding speed, a ``universal'' curve was obtained,
which was valid for all studied exponents in the range $n = 2$ to $8$.

\begin{figure}[tbp]
\includegraphics[width=0.45\textwidth,angle=0.0]{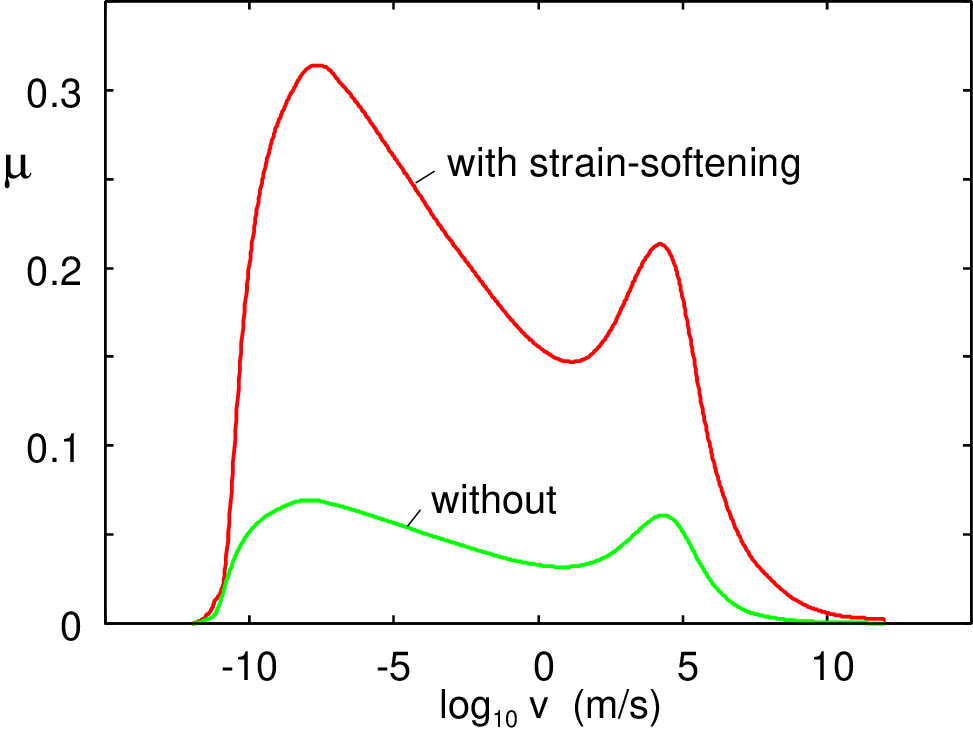}
\caption{\label{1logv.2muROLLING.eps}
Calculated sliding friction coefficient 
with strain softening (red line) and using the low-strain, linear response modulus
(green curve), for the symmetric triangle assuming the measured tip radius $r = 0.1 \ {\rm mm}$.
}
\end{figure}

Rubber filled with reinforcing particles exhibits strong strain softening. 
It is crucial in studies of rolling and sliding friction to account for the rapid decrease in effective modulus with increasing strain. 
This has been emphasized previously in the context of rolling friction \cite{rolling}, but it is equally important for sliding friction. 
In the present study, the strain in the contact region is approximately $0.25$, which is similar to the strain encountered in asperity contacts when a tire is in contact with a road surface. 
Fig. \ref{1logv.2muROLLING.eps} compares the calculated friction using the linear response modulus $E(\omega)$ 
with the result obtained using the effective modulus defined by 
${\rm Re}E_{\rm eff} = \eta_{\rm R} (\epsilon) {\rm Re} E(\omega)$ 
and ${\rm Im}E_{\rm eff} = \eta_{\rm I} (\epsilon) {\rm Im} E(\omega)$. 
The calculated result that includes strain softening yields friction that is approximately five times higher than the result without softening. 
A similarly strong effect of strain softening is observed in rolling friction and in sliding friction on rough surfaces such as road surfaces. 

We have deduced a frictional shear stress $\tau_{\rm f}$ by assuming that there is a contribution to the friction force given by $\tau_{\rm f} A$, where $A$ is the area of real contact. 
If the interface is assumed to be very clean, meaning free from contamination films, the shear stress $\tau_{\rm f}$ may originate from molecular stick-slip processes, as first suggested by Scallamach \cite{Sch} and later studied in more detail by Persson and Volokitin \cite{PV}. 
However, Kl\"uppel \cite{Kl} and Carbone et al. \cite{Carb} have proposed that the adhesive contribution to sliding and rolling friction arises from the propagation of opening and closing cracks at the edges of the contact regions during sliding. 
This mechanism may be an important additional contribution for very soft rubber compounds with smooth surfaces sliding on relatively smooth substrates. 

For the relatively stiff rubber compounds and real surfaces used in most practical applications, such as rubber seals or tires, there is experimental evidence showing that the crack-opening mechanism contributes negligibly to sliding or rolling friction in most cases. 
This is primarily due to the surface roughness, which suppresses or greatly reduces adhesion \cite{Kill}. 
In earlier studies, we showed that rubber compounds exhibiting strong adhesion on very smooth surfaces in the dry state, but no adhesion in water, produce nearly identical sliding friction on rough surfaces in both dry and wet conditions \cite{water}, similar to what Greenwood and Tabor \cite{GT} found.

\vskip 0.3cm
{\bf Declarations}

\vskip 0.2cm
{\bf Ethics approval and consent to participate: }
Not applicable.

\vskip 0.2cm
{\bf Consent for publication: }
Not applicable.

\vskip 0.2cm
{\bf Funding:}
Not applicable.

\vskip 0.2cm
{\bf Authors' contributions: }
All authors contributed equally to the work. All authors read and approved the final manuscript.

\vskip 0.2cm
{\bf Acknowledgements: }
We thank M. Ciavarella for the stimulating discussions that inspired this study.
We thank Shandong Linglong Tire Co., Ltd, for support.

\vskip 0.2cm
{\bf Data availability: }
The data are available from the corresponding author upon reasonable request.

\end{document}